\documentclass{PoS}
\usepackage{wrapfig}
\usepackage{lineno}

\title{First measurements with prototype radio antennas for the IceTop detector array}

\ShortTitle{Prototype radio antennas for IceTop}

\author{
The IceCube Collaboration\footnote{For collaboration list, see PoS(ICRC2019) 1177.}\\
{\itshape \href{http://icecube.wisc.edu/collaboration/authors/icrc19_icecube}{http://icecube.wisc.edu/collaboration/authors/icrc19\_icecube}}\\
E-mail: \email{max.renschler@icecube.wisc.edu}
}

\abstract{

Extending large-scale air-shower arrays with radio antennas can increase the detector's performance, as the radio emission by cosmic-ray air showers provides an additional measurement of the electromagnetic component. Instrumenting the IceCube surface detector IceTop with radio detectors as well as with new particle detectors in a hybrid approach will enhance the measurement and reconstruction accuracy and allow for the characterization of highly inclined air showers. This will enable a better understanding of the atmospheric background for the in-ice neutrino measurements. It also opens the opportunity for new science cases, e.g. the search for PeV gamma rays from the Galactic Center, which is visible from the IceCube site year-round at an inclination of 61$^{\circ}$. Adding to several scintillator particle detectors already running at the South Pole, two prototype radio antennas have been deployed at the IceCube site in January 2019 using the same DAQ system as the scintillators. The antennas serve as a test setup for a future deployment of radio antennas extending the scintillator array planned inside the IceTop footprint. In this proceeding, the antennas considered for deployment and the hybrid DAQ system processing the signals of the particle and radio detectors will be introduced. First measurement results at the South Pole will be presented and future plans for a full hybrid particle and radio detector array inside the IceTop footprint will be shown. 

\vspace{4mm}
{\bfseries Corresponding authors:}
\speaker{Max Renschler}$^{1}$\\
{$^{1}$ \itshape Karlsruhe Institute of Technology, IKP}\\

}

\FullConference{36th International Cosmic Ray Conference -ICRC2019-\\
		July 24th - August 1st, 2019\\
		Madison, WI, U.S.A.}

\begin{document}

\section{Introduction}
IceCube is a cubic-kilometer neutrino detector installed in the ice at the geographic South Pole with the aim to measure and quantify the high-energy neutrino flux \cite{Aartsen:2016nxy}.
The planned upgrade of the IceCube experiment to IceCube-Gen2 foresees an increase of the in-ice detector volume. With this upgrade, an extension of the surface detector IceTop \cite{Gen2} is needed that measures the main background of the IceCube detector consisting of high-energy atmospheric muons induced by cosmic-ray air showers. The current IceTop detector array consists of ice-Cherenkov detectors covering the footprint of the IceCube in-ice detectors of about 1\,km$^2$. As a first step towards a Gen2 surface array, the existing IceTop array is enhanced by adding new detectors consisting of scintillation particle detectors \cite{Scint_Upgrade}, air-Cherenkov telescopes \cite{IceAct} and radio antennas \cite{SPRadio}. This increases the scientific capabilities of the IceTop array and serves as a test setup for a future Gen2 surface array.\\
Adding hybrid particle and radio detectors to IceTop comes with several benefits for the precise measurement of cosmic-ray air showers and also enables new science cases reachable only with a hybrid detector array. Radio emission by air showers has its origin in the separation of charged particles as the air shower propagates through Earth's atmosphere \cite{RadioReviewFrank}. 
The radio emission originates purely from the electromagnetic shower component and is therefore ideal for improving mass composition measurements and the precise tests of hadronic interaction models when measurements are performed together with muon detectors \cite{RadioReviewTim}. Additionally, the measurement of the radio emission of air showers allows for a cross-calibration between different experiments \cite{TunkaCalibration}. \\
Installing surface radio antennas in IceTop enables new science cases, e.g.~the search for PeV gamma rays coming from the Galactic Center (GC). Since the Galactic Center is visible from the IceTop site year-round at an inclination of ~61$^{\circ}$, the site is ideal for searching for inclined air showers generated by PeV gamma rays that might be produced in the Galactic Center \cite{HESS}. The energy threshold of radio to detect PeV gammas can be lowered by using a higher frequency band like 100\,MHz to 190\,MHz than the traditional frequency band of 30\,MHz to 80\,MHz \cite{RadioThreshold}. Radio antennas increase the measurement accuracy of inclined showers significantly since the radio emission is not significantly attenuated during its way through the atmosphere. As radio is sensitive purely to the electromagnetic component, a radio array helps to differentiate between hadronic- and gamma-generated air showers when combined with particle detectors.\\
In January 2019, two prototype antennas were deployed at the South Pole within an already existing scintillation-detector array consisting of seven detectors. Together with the antennas, a first version of a hybrid DAQ system was installed that processes and digitizes the signals of the seven scintillation detectors and the signals of the two radio antennas. This hybrid detector station serves as a test setup for future deployment plans. In addition, measurements of the radio background near the IceCube Laboratory (ICL) were performed to estimate the radio background near the prototype antenna positions.\\   
The final detector layout of the enhanced IceTop array will consist of 32 detector stations with eight scintillation detectors and three radio antennas per station. This adds up to 256 scintillation detectors and 96 radio antennas in an area of 1\,km$^2$ in addition to the original 162 ice-Cherenkov detectors of IceTop \cite{FrankICRC2019}.

\section{The SKALA prototype antenna}\label{sec:antenna}
\begin{figure}
\centering
\includegraphics[height = 0.5\textwidth]{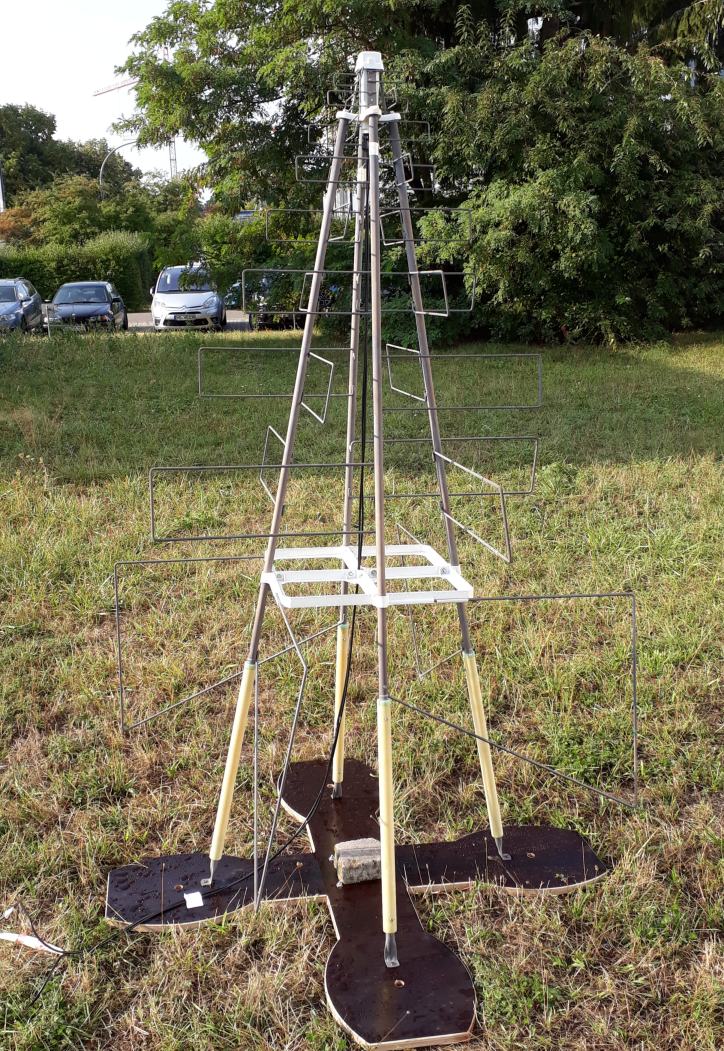}
\hspace{1cm}
\includegraphics[height= 0.5\textwidth]{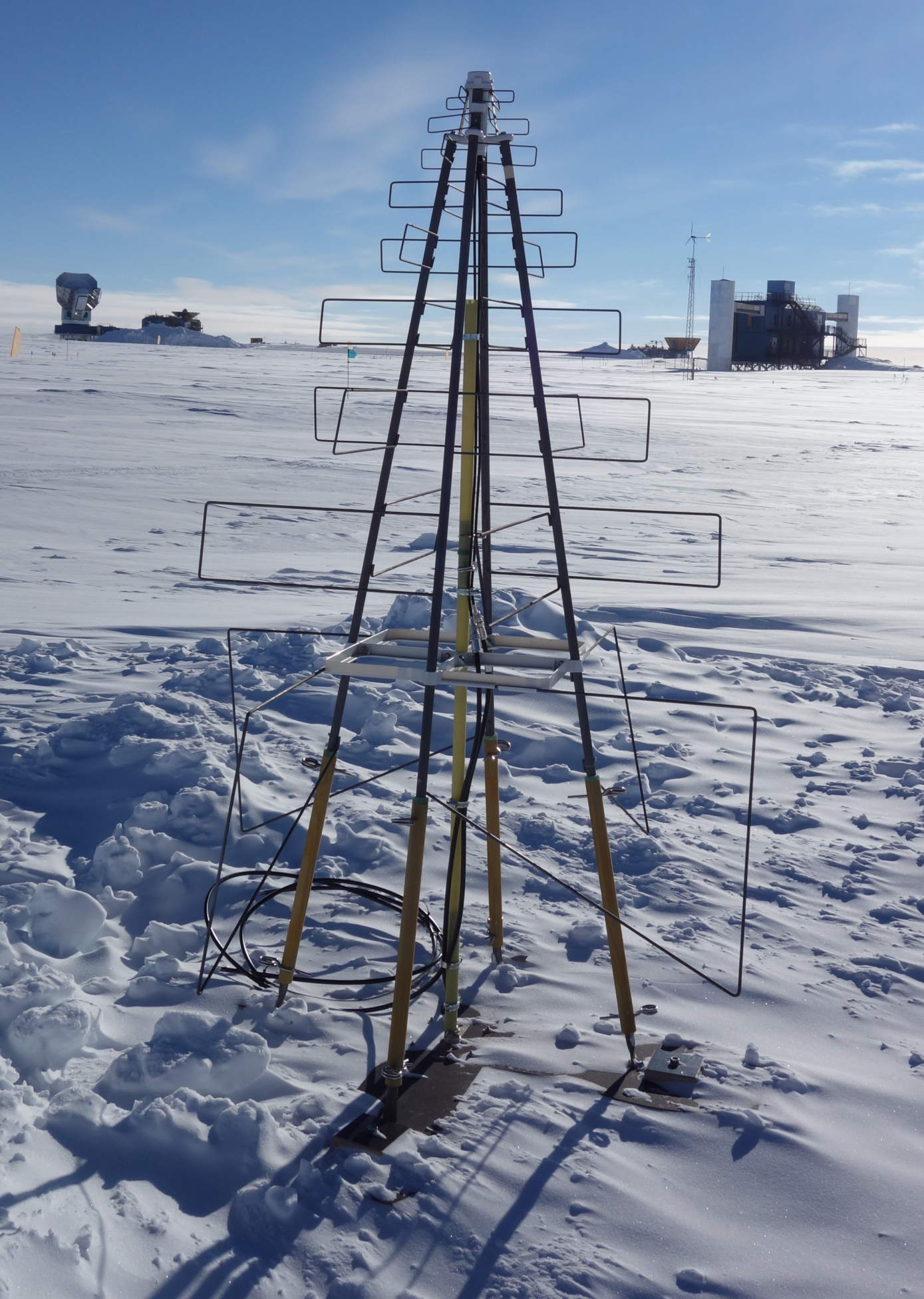}
\caption{SKALA-2 antenna of the South-Pole design with a height of 0.3$\,$m from ground to the lowest lobe. The left picture shows a test setup at the Karlsruhe Institute of Technology (KIT). The right picture shows the deployed antenna at the South Pole.}
\label{pic:SKALA_050}
\end{figure}
A prototype version of an antenna developed by the Square Kilometer Array (SKA) \cite{Skala1} is used. Desirable characteristics of the antenna include the gain of the antenna for inclined showers down to an inclination of 61$^{\circ}$, the low internal electronic noise of about 30\,K and the wide frequency bandwidth of 70\,MHz to 350\,MHz.\\
The antenna structure and the antenna itself have to withstand the harsh environmental conditions at the South Pole. For this, a new antenna structure made out of glass-fiber reinforced plastic (GFK) material and plywood has been designed, and critical antenna parts like plastic mountings have been stress-tested at low temperatures.\\
Pictures of the antenna are shown in Fig. \ref{pic:SKALA_050}. The antenna arms are extended with GFK poles leading to a small metal foot that is attached to a plywood mounting structure. The mounting structure is attached to the ground with four GFK snow spikes with a length of 1\,m each. A middle pole serves as an additional stabilization and as a cable guide from the low-noise amplifier (LNA) on the top of the antenna. Future designs of the antenna structure will enable the possibility to lift the antenna when the snow level rises during years of operation.\\ 
The antenna structure has been designed in three different heights of 0.3\,m (shown in Fig. \ref{pic:SKALA_050}), 0.7\,m and 1.7\,m from ground level to the lowest antenna lobe resulting in a total height of the antenna of 1.8\,m, 2.3\,m and 3.3\,m. The shortest and tallest antenna designs were deployed at the South Pole in January 2019.\\
Prior to the deployment at the South Pole, the LNA of the antenna was tested and characterized down to temperatures of -70$^{\circ}$C. For this, the LNA response to an input signal generated by a calibration source was measured while temperature cycling the LNA from 20$^{\circ}$C to -70$^{\circ}$C. 
\begin{figure}
\centering
\includegraphics[width = 0.85\textwidth]{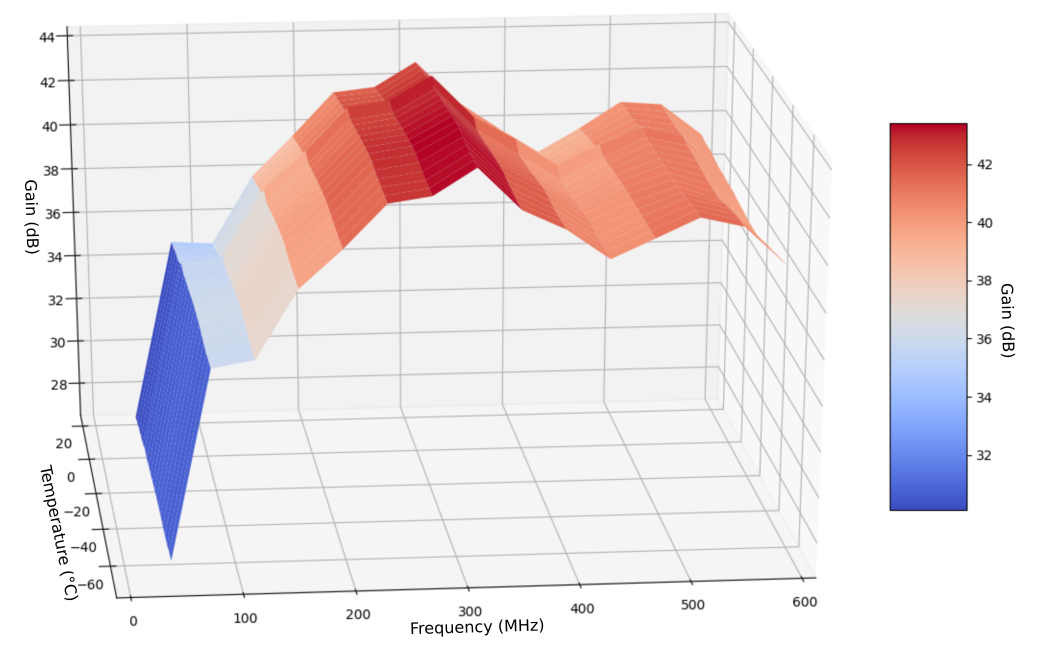}
\caption{Measurement of the LNA gain. The gain is measured in a temperature range of 20$^{\circ}$C to -70$^{\circ}$C and a frequency range of 50\,MHz to 600\,MHz. As input, a 50\,$\Omega$ calibration source has been used.}
\label{pic:LNA_gain}
\end{figure}
In Fig. \ref{pic:LNA_gain} the results of the LNA gain measurement in a temperature range of 20$^{\circ}$C to -70$^{\circ}$C and a frequency range up to 600\,MHz are shown. The LNA gain stays constant within about 2\,dB throughout the whole temperature range, allowing for stable operation in the cold and temperature-dynamic environment of the South Pole. 

\section{Hybrid particle and radio detector DAQ}\label{sec:DAQ}
The scintillation detectors and the radio antennas share a common DAQ system which is based on the Transportable Array for eXtremely large area Instrumentation studies (TAXI) \cite{TAXI}. To be able to digitize scintillation and radio detector signals with TAXI, the original design of TAXI has been adapted. To add essential functions for the radio-signal processing, a radio front-end electronic board has been designed and embedded into the system.\\
\begin{figure}
\centering
\includegraphics[width = 0.85\textwidth]{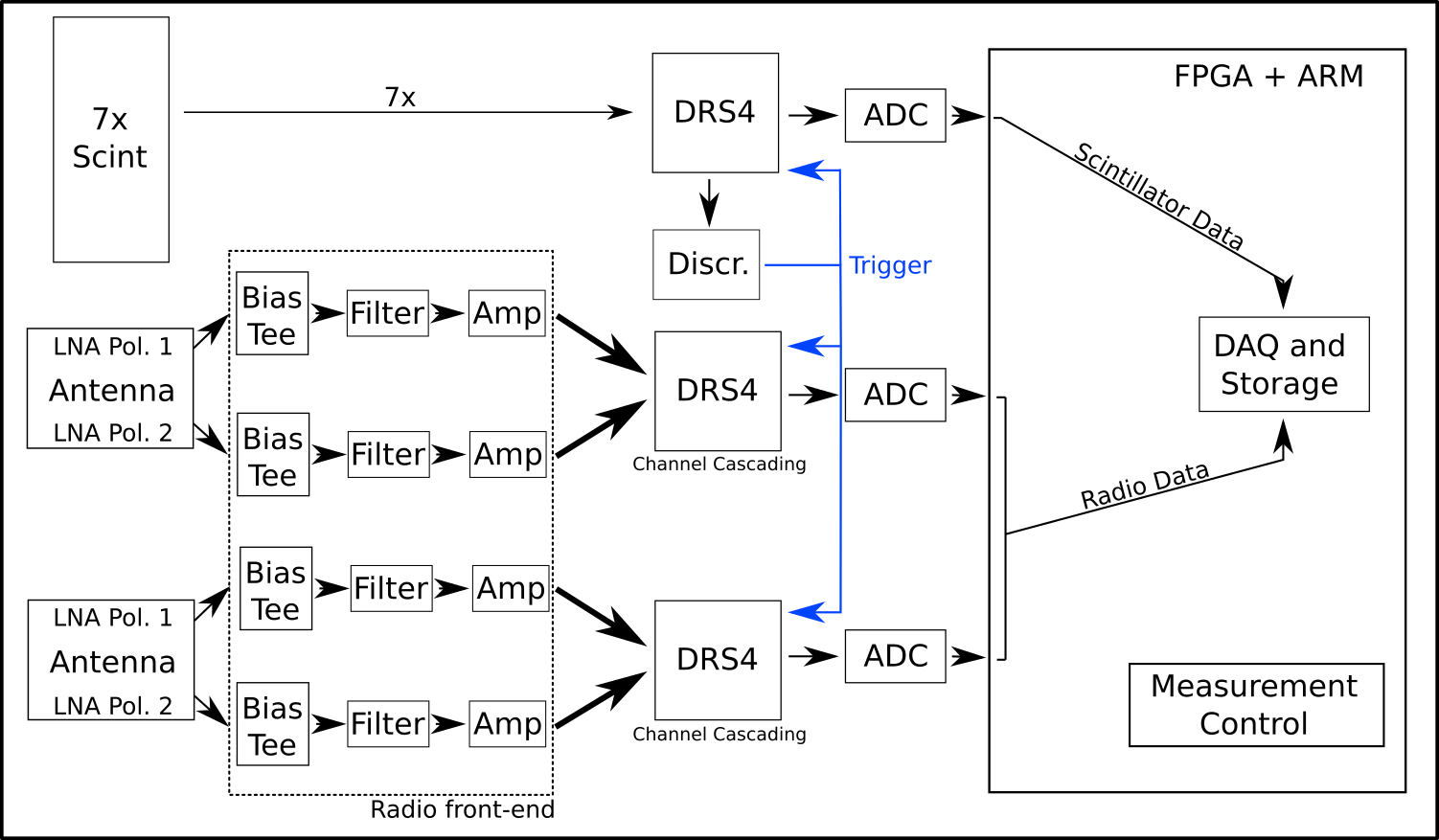}
\caption{Block diagram of the hybrid TAXI DAQ system.}
\label{pic:RadioTAXI}
\end{figure}
In Fig. \ref{pic:RadioTAXI} a block diagram of the hybrid DAQ system is shown. The TAXI DAQ samples the incoming analog signal with DRS4 sampling chips. Each DRS4 chip has eight input channels with 1024 sampling cells each. The sampling frequency of the DRS4 chips is adjustable from 0.7\,GHz to 5\,GHz. A sampling rate of 1.0\,GHz is used. After a trigger event, all channels of each DRS4 are read out through an eight-channel \textit{LTM9007IY-14} analog-to-digital converter (ADC) with 14-bit sampling depth. The trigger is realized with a comparator as a signal-over-threshold trigger with adjustable threshold voltage. This trigger is based on the scintillation detectors and requires the coincidence of a variable number of triggered scintillation detectors within a time window of 180\,ns. In future designs a trigger based on the radio signal will also be available that enables the option of a radio self-trigger.\\
The TAXI DAQ is designed to house three DRS4 sampling chips of which one samples the incoming scintillation detector signals. The remaining two DRS4 chips digitize the signals of two radio antennas with two polarizations each. For the radio-signal sampling the DRS4 chips are operated in channel-cascading mode which reduces the available DRS4 channels to two per DRS4 but increases the number of sampling cells to 4096 for each channel resulting in a radio-trace length of about 4\,$\mu$s.\\
Radio-specific components are placed on a radio front-end board before the TAXI inputs. The radio front-end electronics consist of a bias tee (\textit{JEBT-4R2G+}), a protection against electrostatic discharge (ESD) reaching the electronics over the antenna signal cable, a high-pass filter (\textit{SXHP-48+}) and a low-pass filter (\textit{ULP-340+}) which limit the measured signal bandwidth to about 50--350\,MHz. To enable the DRS4 channel cascading, the antenna signal has to be fanned out into four DRS4 inputs. For this, the filtered radio signal is passively fanned out into the inputs of four parallel amplifiers stages that amplify the signal by 10\,dB and convert the single-ended input signal to a differential output signal. The gain of the amplifier stages is chosen to counteract the attenuation from the signal cable and prior electronic components.
The fanned-out and amplified signals are transferred via CAT-7 cables with RJ45 connectors to an adapter board that connects the signals to the respective TAXI inputs.

\section{Deployment at the South Pole}\label{sec:SP}
%
\begin{figure}
\centering
\includegraphics[width = 0.4\textwidth]{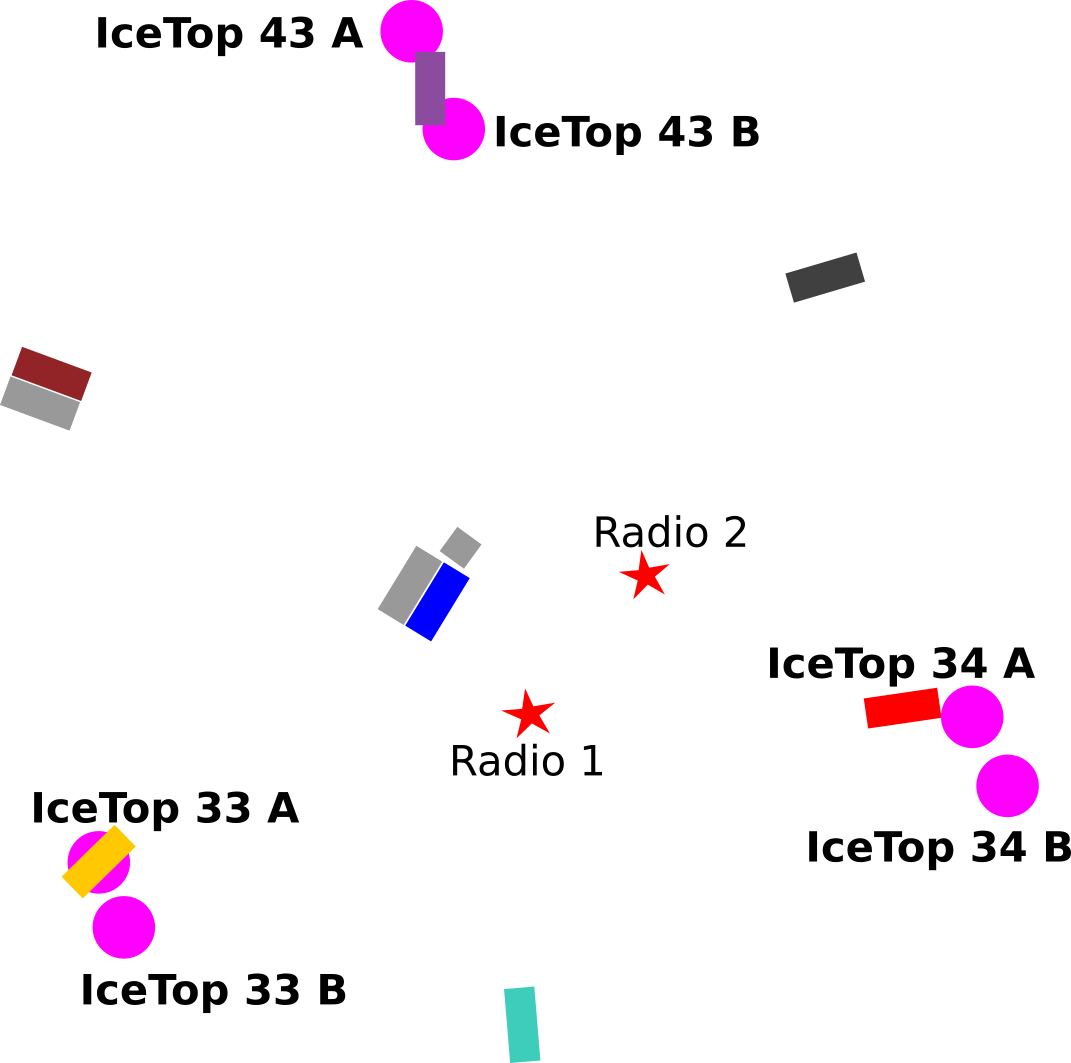}
\caption{Not-to-scale layout of the deployed scintillation and radio detectors with scintillation detectors as rectangles, IceTop tanks as circles and radio antennas as stars (January 2019).}
\label{pic:detector_layout}
\end{figure}
%
In January 2019, the existing scintillation detector array was extended by two antennas and a new DAQ system capable of processing the scintillation and radio detector signals (Fig.~\ref{pic:RadioTAXI}). In Fig. \ref{pic:detector_layout}, a sketch of the detector layout is shown. The antennas are located in the middle of the array, next to the central scintillation detectors and the DAQ electronics. The antenna polarizations aim to possible sources of radio background like the ICL, the DAQ electronics and other detectors. The goal of this arrangement is to measure the radio background in the field and to identify strong noise sources producing radio-frequency interference (RFI) peaks in the targeted frequency bandwidth.\\
\begin{figure}
\centering
\includegraphics[width = \textwidth]{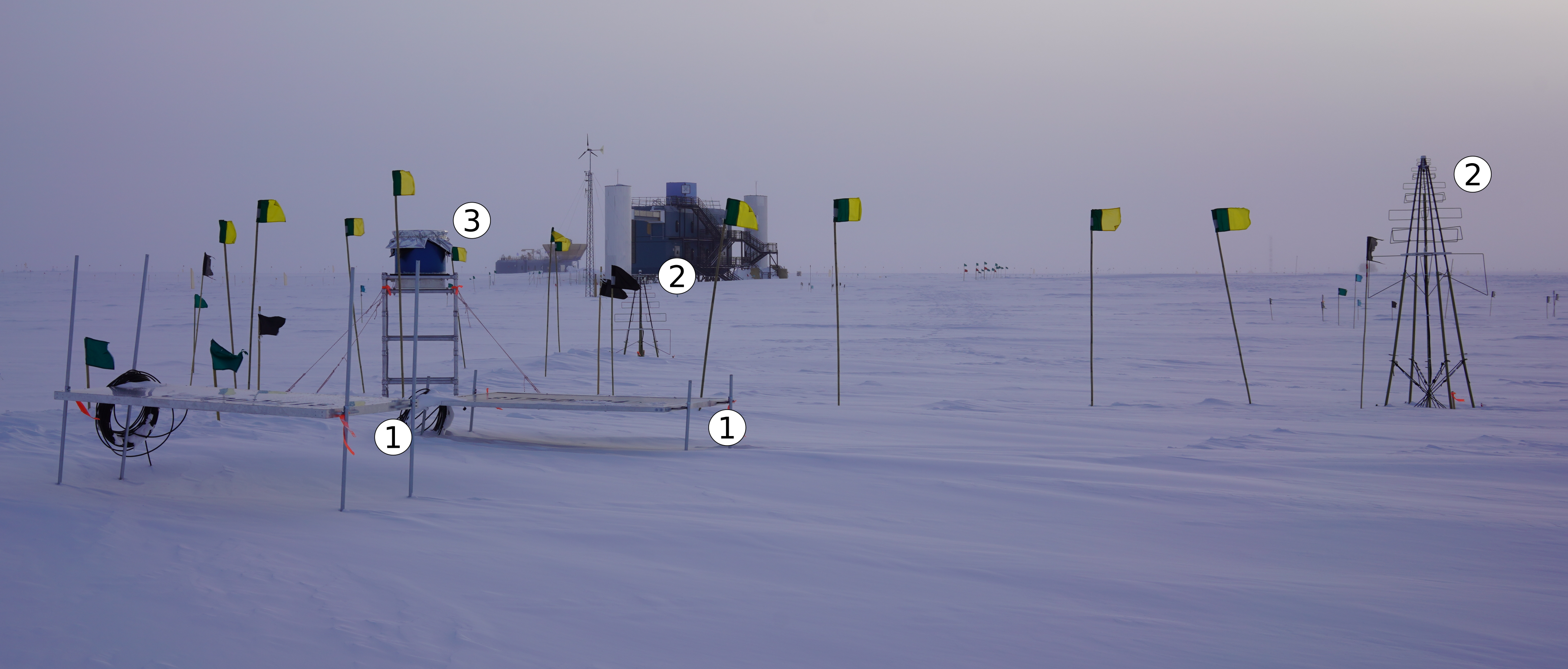}
\caption{Detector site with scintillation detectors (1), radio antennas (2) and elevated IceAct camera (3).}
\label{pic:SP_detector_station}
\end{figure}
In Fig. \ref{pic:SP_detector_station}, a photo of the detector setup at the South Pole in January 2019 is shown. The picture shows the two deployed antennas together with the central scintillation detectors and the IceAct camera. The antenna mechanical structure was found to be fast and easy to deploy, an important criterion for a future deployment of 96 antennas in the footprint of IceTop. The two antennas were deployed in a period of two days with a team of two people. The deployment may be sped up to two to three antennas per day with an experienced team. 

\section{First measurement results}\label{sec:Measurements}
Parallel to the deployment of the antennas and the hybrid DAQ electronics, background measurements near the ICL were performed to estimate the radio background prevalent near the detectors. For these measurements, the 2.3\,m-high antenna has been used. The antenna was connected with a 30\,m cable to a commercial DAQ system (CAEN \textit{DT5730}) located inside of the ICL. The distance between the ICL and the antenna was roughly 25\,m. The nearby scintillation detectors and the IceAct camera were powered off.\\
\begin{figure}
\centering
\includegraphics[width = 0.9\textwidth]{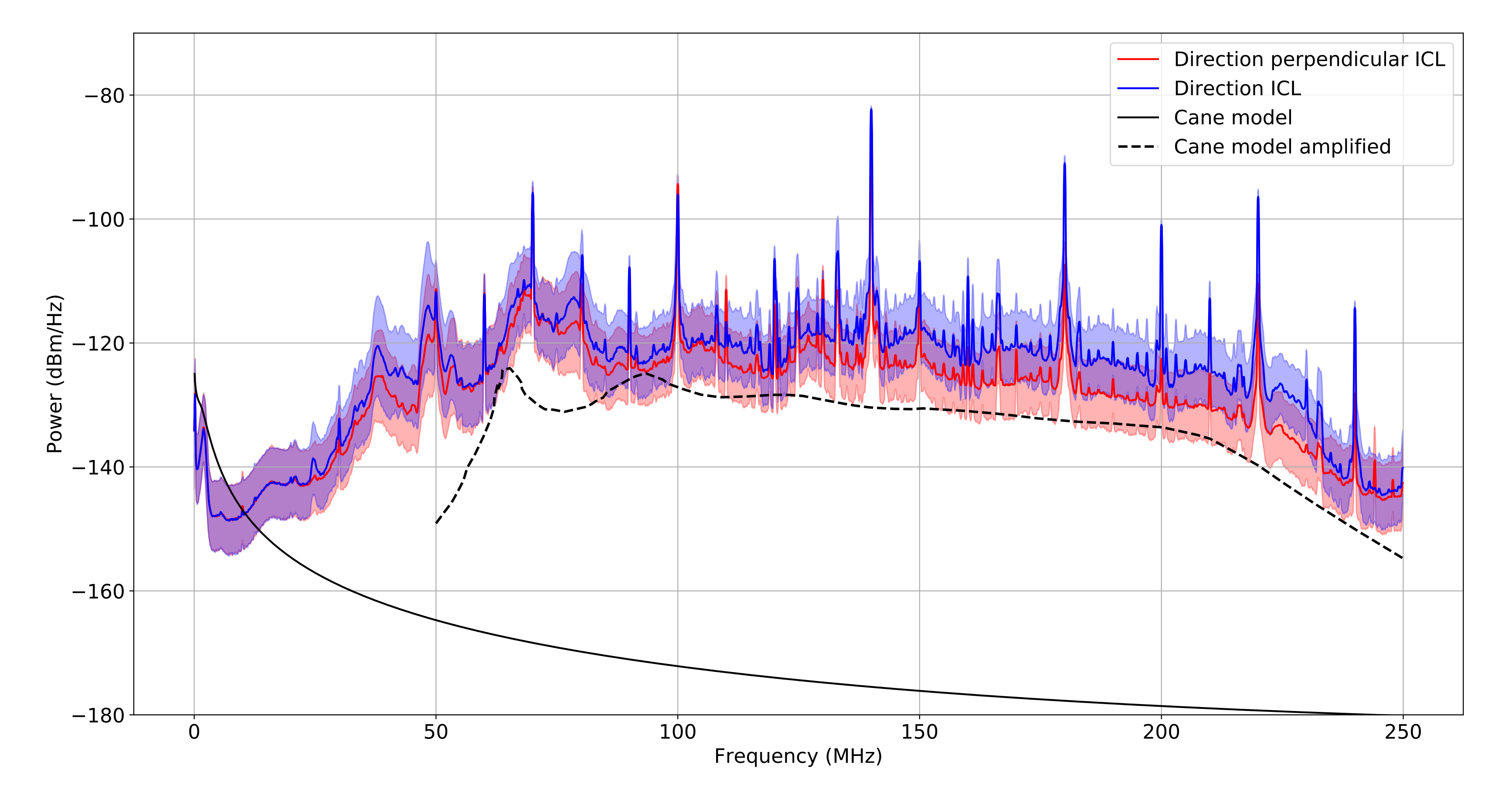}
\caption{Measurement of the background frequency spectrum near the ICL. The black dashed line shows the expected background noise originating from the Galactic Center and other sources in the sky.}
\label{pic:SP_background_measurement}
\end{figure}
In Fig. \ref{pic:SP_background_measurement},~the average frequency spectrum of two one-hour measurements in the direction of the ICL (blue) and perpendicular to the ICL (red) are shown. Comparing the two measurements, the expected behaviour of an increase in noise when aiming the antenna polarization towards the ICL is visible. Both spectra show unexpectedly high RFI peaks. These peaks become visible at 50\,MHz and reappear every 10\,MHz. The source of this RFI noise is unknown. With available information from the Antarctic Service Contractor (ASC), sources such as radio communication or beacons can be excluded. Other experiments performing background studies like ARA \cite{ARA1, Ara2} and previous field measurements \cite{RASTA_Background} did not observe these RFI signals. Since the presented measurements have been performed at a distance of about 25\,m to the ICL, which houses a workshop with electronic equipment on the first floor, a noise source inside of the ICL is possible. Measurements with the deployed antenna farther away from the ICL will help understand this radio background. \\
Fig. \ref{pic:SP_background_measurement} also includes the prediction of the sky radio background as a solid black line and the expected amplified sky radio background as a dashed black line. The amplified sky radio background includes the gain of the antenna, the LNA, the low-pass filter and the cable together with a constant thermal noise of 30\,K coming from the antenna and the LNA. As LNA response, the simulated LNA gain from E. de Lera Acedo et al. \cite{SKALAV2_LNA} has been used that stops at a frequency of 50\,MHz. The overall measured frequency spectra is higher than the expectation of the sky noise. For the sky noise, the Cane model \cite{Cane_model} has been used which predicts an average sky noise originating from the whole sky including the Galactic Center. Due to this averaging over the whole sky, a slight deviation of the expectation and the measurement is expected. The origin of the discrepancy between the measurement and the prediction is unclear. The deviation could come from an additional thermal noise contribution which has not been included in the amplified background estimation. Measurements with the deployed antennas will give additional information about the overall noise situation.

\section{Conclusion and Outlook}
Two prototype antennas together with a hybrid DAQ system were deployed at the South Pole in January 2019. Tests of the antenna mechanics as well as the antenna LNA demonstrate the ability of the antenna to withstand very low temperatures. First background measurements near the detector site and next to the ICL show a frequency spectrum higher than the prediction of the sky radio background and with a contamination of RFI signals of unknown origin. Work on investigating the radio background and improving the radio front-end electronics as well as the antenna mechanics is ongoing, with an aim to deploy an improved test station at the South Pole in the 2019--20 season.\\
These test deployments are preparing for a proposed final detector array consisting of in total 256 scintillation detectors and 96 radio antennas. The array will be subdivided in stations of eight scintillation detectors and three radio antennas wthat share one DAQ system \cite{ScintsICRC2019}. 

\bibliographystyle{ICRC}
\bibliography{references}

\providecommand{\href}[2]{#2}\begingroup\raggedright\begin{thebibliography}{10}

\bibitem{Aartsen:2016nxy}
{\bf IceCube} Collaboration, M.~G. Aartsen et~al., {\em JINST} {\bf 12} (2017)
  P03012.

\bibitem{Gen2}
{\bf IceCube} Collaboration, M.~G. Aartsen et~al.,
  \href{http://arxiv.org/abs/1412.5106}{{\tt arXiv:1412.5106}}.

\bibitem{Scint_Upgrade}
{\bf IceCube-Gen2} Collaboration, S.~Kunwar, T.~Huber, J.~Kelley, and D.~Tosi,
  \pos{PoS(ICRC2017)401} (2018).

\bibitem{IceAct}
{\bf IceCube Gen2} Collaboration, J.~Auffenberg,  \pos{PoS(ICRC2017)1055}
  (2018).

\bibitem{SPRadio}
F.~G. {Schr\"oder}, {\em EPJ Web Conf.} {\bf 208} (2019) 15001.

\bibitem{RadioReviewFrank}
F.~G. {Schr\"oder}, {\em Prog. Part. Nucl. Phys.} {\bf 93} (2017) 1--68.

\bibitem{RadioReviewTim}
T.~Huege, {\em Phys. Rept.} {\bf 620} (2016) 1--52.

\bibitem{TunkaCalibration}
{\bf Tunka-Rex, LOPES} Collaboration, W.~D. Apel et~al., {\em Phys. Lett.} {\bf
  B763} (2016) 179--185.

\bibitem{HESS}
{\bf HESS} Collaboration, {\em Nature} {\bf 531} (2016) 476 -- 479.

\bibitem{RadioThreshold}
A.~Balagopal~V., A.~Haungs, T.~Huege, and F.~G. {Schr\"oder}, {\em Eur. Phys.
  J.} {\bf C78} (2018) 111.

\bibitem{FrankICRC2019}
{\bf IceCube} Collaboration, F.~G. {Schr\"oder},  \pos{PoS(ICRC2019)418,these
  proceedings} (2019).

\bibitem{Skala1}
E.~de~Lera~Acedo, N.~Razavi-Ghods, N.~Troop, N.~Drought, and A.~J. Faulkner,
  {\em Experimental Astronomy} {\bf 39} (Oct, 2015) 567--594.

\bibitem{TAXI}
T.~Karg, A.~Haungs, M.~Kleifges, R.~Nahnhauer, and K.~H. Sulanke, {\em {6th
  International Workshop on Acoustic and Radio EeV Neutrino Detection
  Activities (ARENA 2014) Annapolis, MD, June 9-12, 2014}} (2014).

\bibitem{ARA1}
{\bf ARA} Collaboration, P.~{Allison}, J.~{Auffenberg}, R.~{Bard}, J.~J.
  {Beatty}, D.~Z. {Besson}, S.~{B{\"o}ser}, C.~{Chen}, P.~{Chen}, and
  A.~{Connolly}, {\em Astroparticle Physics} {\bf 35} (Feb, 2012) 457--477.

\bibitem{Ara2}
{\textbf{ARA, IceCube} Collaborations, S. {B{\"o}ser}}, {\em AIP Conference
  Proceedings} {\bf 1535} (2013) 116--120.

\bibitem{RASTA_Background}
J.~Auffenberg, T.~Gaisser, K.~Helbing, T.~Huege, T.~Karg, and A.~Karle, {\em
  Nucl. Instrum. Meth.} {\bf A604} (2009) S53--S56.

\bibitem{SKALAV2_LNA}
E.~{de Lera Acedo}, N.~{Drought}, B.~{Wakley}, and A.~{Faulkner}, {\em 2015
  International Conference on Electromagnetics in Advanced Applications
  (ICEAA)} (Sep., 2015) 839--843.

\bibitem{Cane_model}
H.~V. {Cane}, {\em Monthly Notices of the Royal Astronomical Society} {\bf 189}
  (Nov., 1979) 465--478.

\bibitem{ScintsICRC2019}
{\bf IceCube} Collaboration, M.~Kauer,  \pos{PoS(ICRC2019)309,these
  proceedings} (2019).

\end{thebibliography}\endgroup


\providecommand{\href}[2]{#2}\begingroup\raggedright\endgroup

\end{document}